\documentclass[aps,prl,twocolumn,]{revtex4}
\usepackage{amsmath}
\usepackage{amsfonts}
\usepackage{amssymb}
\usepackage{bm}
\usepackage{color}
\usepackage{graphicx}

\def\uk{{\bm u}({\bm k})}

\def\upp{u^+({\bm p})}
\def\uqm{u^-({\bm q})}
\def\uqm{u^-({\bm q})}

\def\ukp{u^+({\bm k})}
\def\ukm{u^-({\bm k})}
\def\bk{\bm k}
\def\bp{\bm p}
\def\br{\bm r}
\def\bq{\bm q}
\def\bv{\bm v}
\def\bx{\bm x}
\def\bw{\bm w}
\def\hm{{\bm h}^-({\bm k})}
\def\hp{{\bm h}^+({\bm k})}

\begin{document}
\title{Inverse energy cascade in three-dimensional isotropic
  turbulence} \author{Luca Biferale$^{1}$, Stefano Musacchio$^2$ \&
  Federico Toschi$^{3}$} \affiliation{$^1$Dept. Physics \& INFN,
  U. Tor Vergata, Via della Ricerca Scientifica 1, 00133 Rome, Italy.}
\affiliation{$^2$ CNRS, Lab. J.A. Dieudonn\'e UMR 6621, Parc Valrose,
  06108 Nice, France.}
\affiliation{$^3$Dept. Physics \& Dept. Mathematics and Computer
  Science Eindhoven University of Technology, 5600 MB Eindhoven The
  Netherlands \& CNR-IAC, Via dei Taurini 19, 00185 Rome, Italy.}
\date{\today}

\begin{abstract}
  In turbulent flows kinetic energy is spread by nonlinear
  interactions over a broad range of scales.  Energy transfer may
  proceed either toward small scales or in the reverse direction.  The
  latter case is peculiar of two-dimensional (2D) flows.
  Interestingly, a reversal of the energy flux is observed also in
  three-dimensional (3D) geophysical flows under rotation and/or
  confined in thin layers.  The question is whether this phenomenon is
  enforced solely by external anisotropic mechanisms or it is
  intimately embedded in the Navier-Stokes (NS) equations.  Here we
  show that an inverse energy cascade occurs also in 3D isotropic
  flow.  The flow is obtained from a suitable surgery of the NS
  equations, keeping only triadic interactions between sign-defined
  helical modes, preserving homogeneity and isotropy and breaking
  reflection invariance.  Our findings highlight the role played by
  helicity in the energy transfer process and show that both 2D and 3D
  properties naturally coexist in all flows in nature.
\end{abstract}
\maketitle
Inviscid invariants of the NS equations are crucial in determining the
direction of the turbulent energy transfer \cite{frisch}.  In some
cases, as for fully isotropic and homogeneous turbulence in 2D the
presence of two positive-defined invariants (energy and enstrophy)
does not allow a stationary transfer of both quantities, neither to
small nor to large scales \cite{kraichnan_2d}.  In presence of two
fluxes, they must necessarily flow in opposite directions
\cite{tabeling,ecke,boffi_2d,2dreview,prl_massimo} and this remains
true even for turbulent systems in non-integer dimensions obtained by
fractal Fourier decimation \cite{fractal}.  The fluid equations
possess two inviscid invariants also in 3D: energy and helicity
(i.e. the scalar product of velocity and vorticity).  The inviscid
conservation of helicity was discovered relatively
recently~\cite{moffat,brissaud}.  At variance with energy, helicity is
not positive defined.  This allows for a simultaneous small-scale
transfer of energy and helicity, as confirmed by results of two-point
closures~\cite{brissaud,lesieur,waleffe} and direct numerical
simulations~\cite{pouquet_dns,chen}.  Nevertheless, a reversal of the
flux of energy has been observed in geophysical flows subject to earth
rotation \cite{smith96,pouquet} as well as in shallow fluid
layers~\cite{gage-nastrom,Lauthnschaleger,smith99,cmv2010,xia}.  In
both cases, this phenomenon is accompanied by strong anisotropic
effects and by a substantial two-dimensionalization of the flow,
induced either by the rotation or by the effects of
confinement. Moreover, rotations injects fluctuations in the helical
sector while a perfect two-dimensional flow has vanishing {\it
  point-wise} helicity, being vorticity always ortogonal to velocity.
\begin{figure*}[!th]
  \centerline{\includegraphics[scale=0.5,draft=false]{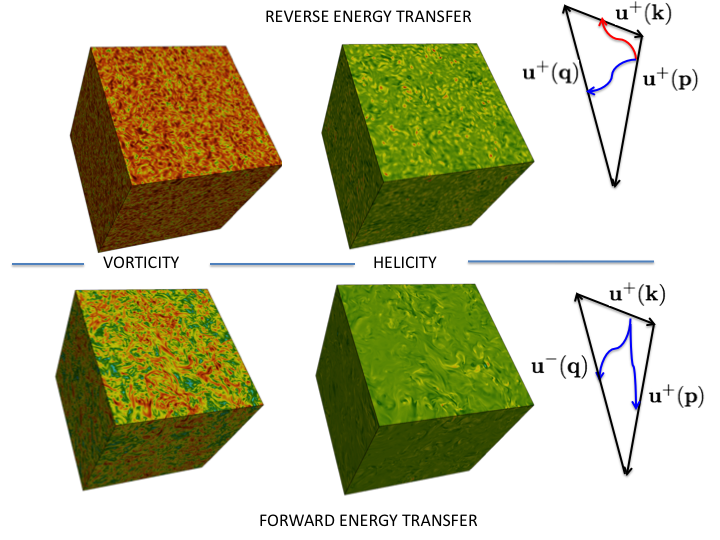}}
  \caption{Comparison between helicity and vorticity fields for normal
    NS turbulence (bottom row) and for inverse cascade 3D turbulence
    (top row).  We also represent a pictorial scheme of triads
    responsible for the inverse cascade regime as suggested in
    \cite{waleffe}. Inverse 3D cascade is possible with triads where
    all helical components have the same sign. In this case the middle
    wave number (here $\upp$) transfers energy also to smaller
    wavenumbers ($\ukp$), at difference from what happens for triads
    with different helical components where it is always the smallest
    wavenumber $\ukp$ that transfer energy to higher modes
    $\uqm,\upp$.}
  \label{fig:1}
\end{figure*}
The role played by helicity in the energy transfer mechanism of 3D
flows has attracted a broad scientific interest (see,
e.g.,~\cite{chen} and reference therein).  Dynamical systems have been
developed to study in details energy and helicity transfer at high
Reynolds numbers~\cite{biferale_hel,ditlevsen}.  Further, speculation
connecting the existence of intermittent burst in the energy cascade
induced by ``local'' helicity blocking mechanism have been
proposed~\cite{biferale_hel}. Despite these important contributions,
the understanding of the phenomenology of helicity remains
``mysterious'', as summarized in the conclusion of a recent
state-of-the-art numerical study~\cite{chen}.
Here we present theoretical and numerical evidences of a new
phenomenon induced by helicity conservation: a statistically
stationary {\it backward} energy transfer can be sustained even in 3D
fully isotropic turbulence.

The starting point of our analysis is the well-known helical
decomposition \cite{waleffe} of the velocity field $\bv(\bx)$,
expanded in Fourier series, $\uk$:
\begin{equation}
  \uk  = \ukp \hp + \ukm \hm
\end{equation}
where $ {\bm h}^\pm$ are the eigenvectors of the curl operator $i {\bm
  k} \times {\bm h}^\pm = \pm k {\bm h}^\pm$.  In particular, we
choose $ {\bm h}^\pm =\hat{\bm \nu} \times \hat{\bm k} \pm i \hat{\bm
  \nu}$, where $ \hat{\bm \nu}$ is an arbitrary versor orthogonal to
${\bm k}$ which satisfies the relation $\hat{\bm \nu}({\bm k}) = -
\hat{\bm \nu}(-{\bm k})$ (necessary to ensure the reality of the
velocity field).  Such requirement is satisfied e.g. by the choice
$\hat{\bm \nu} = {\bm z} \times {\bm k} /|| {\bm z} \times {\bm k}
||$, with ${\bm z}$ an arbitrary vector.  In terms of this {\it exact}
decomposition of each Fourier mode energy, $E = \int d^3 x
|\bv(\bx)|^2$, and helicity, $H = \int d^3 x \bv \cdot \bw$, where
$\bw$ is the vorticity, are written as:
\begin{equation}
  \begin{cases}
    E = \sum_{\bk} |\ukp|^2 + |\ukm|^2; \label{eq:E}\\
    H = \sum_{\bk} k(|\ukp|^2 - |\ukm|^2).
  \end{cases}
\end{equation}
Similarly, the non-linear term of the NS equations can be exactly
decomposed in 4 independent classes of triadic interactions,
determined by the helical content of the complex amplitudes, ${\bm
  u}^{s_k}(\bk)$ with $s_k = \pm $ (see \cite{waleffe}).  Among three
generic interacting modes ${\bm u}^{s_k}(\bk),{\bm u}^{s_p}(\bp), {\bm
  u}^{s_q}(\bq)$, one can identify 8 different helical combinations
$(s_k=\pm,s_p=\pm,s_q=\pm)$.  Among them, only four are independent
because of the symmetry that allows to change all signs of helicity
simultaneously.
\begin{figure}[!th]
  \centerline{
    \includegraphics[scale=0.75,draft=false]{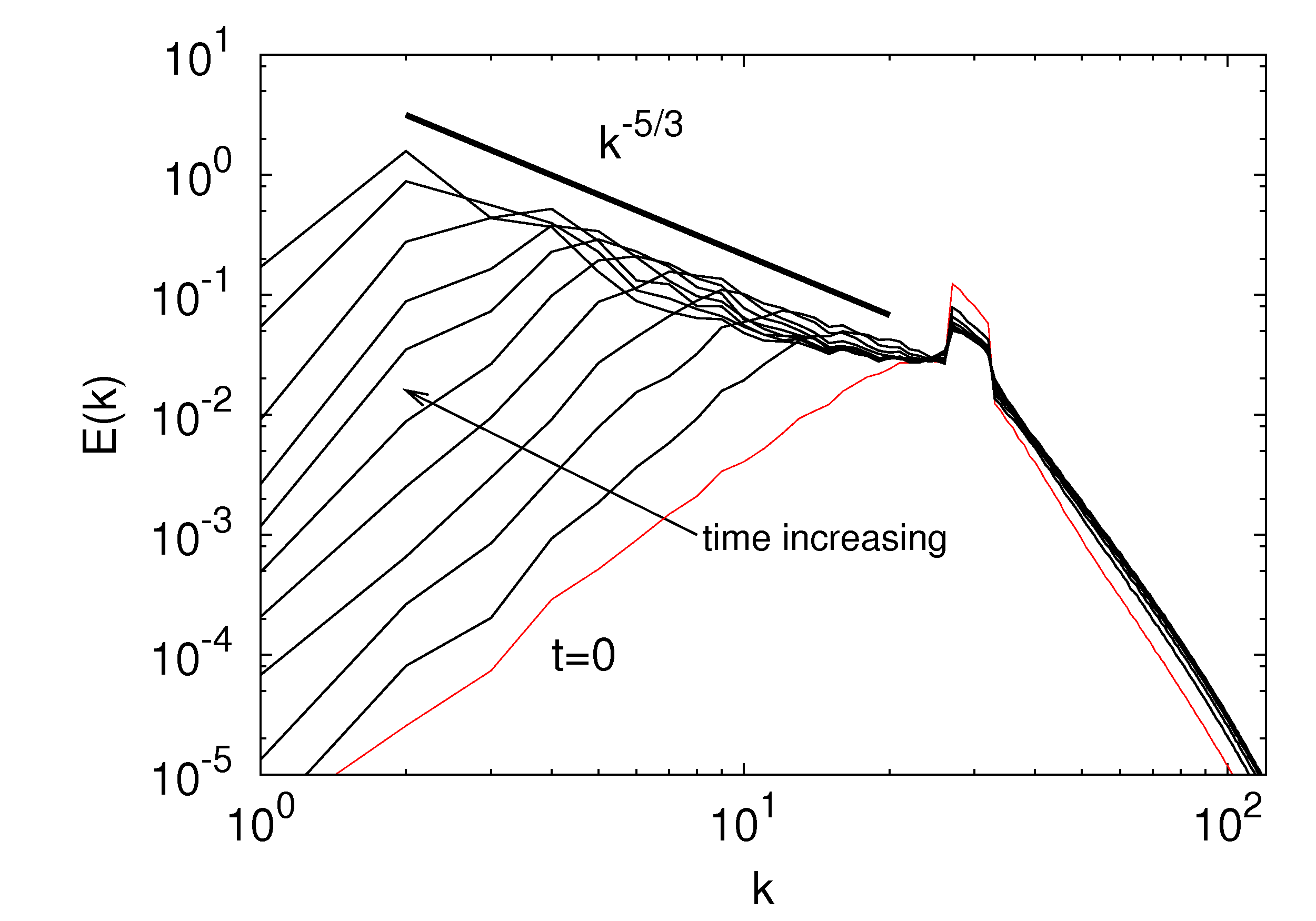}}
  \caption{(color online) Non-stationary spectrum in the inverse
    energy cascade regime.  Red, dashed line represent $k^{-5/3}$
    slope.}
  \label{fig:2}
\end{figure}
Let us now consider the dynamics of an incompressible flow ${\bm
  \nabla} \cdot {\bm v} = 0$, which is determined by a decimated NS
equation in which all interactions between modes have been switched
off except for those with a well defined sign of helicity,
e.g. positive $(s_k=+,s_p=+,s_q=+)$.  We define the projector on
positive/negative helicity states as
\begin{equation}
  \label{eq:proj}
  {\mathcal P}^\pm \equiv 
  \frac {{\bm h}^\pm \otimes \overline{{\bm h}^{\pm}}} 
  {\overline{{\bm h}^\pm} \cdot {\bm h}^\pm}
\end{equation}
where $\overline{\cdot}$ stands for complex conjugate.  Then we
project the velocity field into its positive helicity component:
\begin{equation} 
  \label{eq:projv}
  {\bv}^+(\bx) \equiv  \sum_{\bk} e^{i{\bm k}\bx}{\mathcal P}^+ {\uk};
\end{equation}
and we consider the decimated NS equations:
\begin{equation} 
  \label{eq:ns+++}
  \partial_t {\bm v}^+ = (- {\bm v}^+ \cdot {\bm \nabla} {\bm v}^+ -{\bm \nabla} p )^+ 
  +\nu \Delta {\bm v}^+ + {\bm f}^+ 
\end{equation}
where $\nu$ is the viscosity, $p$ is the pressure and ${\bm f}$ is the
external forcing stirring the fluid around a wavevector $k_f$.  The
non-linear term and the forcing are projected on the positive helicity
states with the same procedure followed for the velocity field
(\ref{eq:projv}).  The resulting system has two positive definite
invariants, see Eq.~(\ref{eq:E}), the energy and the helicity, $ H =
\sum_{\bk} k |\ukp|^2$, and contains only interactions between
positive helicity modes.  Helicity becomes a coercitive quantity: the
decimated NS equations cannot sustain a simultaneous forward cascade
of energy and helicity, for the same arguments which forbid the
existence of a simultaneous forward cascade of energy and enstrophy in
2D turbulence~\cite{kraichnan_2d,waleffe}.  Therefore, the dynamics of
Eq.~(\ref{eq:ns+++}) should display a double cascade phenomenology,
characterized by an inverse energy cascade with Kolmogorov spectrum
$E(k) \sim k^{-5/3}$ for $k \ll k_f$, and a direct helicity cascade
with a $k^{-7/3}$ spectrum for $k \gg k_f$.  It is interesting to note
that, at variance with usual 3D NS dynamics, such flow should not
display dissipative anomaly for kinetic energy, i.e. energy
dissipation should vanish in the limit $\nu \to 0$.  Indeed, the
direct helicity cascade carries also a residual, non-constant flux of
kinetic energy toward small scales which decays as $k^{-1}$ and
therefore vanishes in the high Reynolds number limit. As a
consequence, one may speculate that the decimated NS equations posses
a less singular spatio-temporal evolution.
Numerical simulations has been performed with a fully-dealiased,
pseudo-spectral code at resolution $512^3$ on a triply periodic cubic
domain of size $L=2\pi$.  The flow is sustained by a random Gaussian
forcing, with $\langle f_i(\bk,t) f_j(\bq,t') \rangle = F(k)
\delta(\bk-\bq) \delta(t-t') Q_{i,j}(\bk)$, where $Q_{ij}(\bk)$ is a
projector assuring incompressibility and $F(k)$ has support only in
the high wavenumber range $ |k| \in [k_{min}=25:k_{max}=32]$.

A visual inspection of the helicity and vorticity fields offered in
Fig.~\ref{fig:1} shows the differences between the forward cascade,
which develops in standard 3D NS equations forced at large scale, and
the novel 3D inverse cascade regime obtained from the decimated NS
equation~(\ref{eq:ns+++}) forced at small scales (see
Fig.~\ref{fig:1}).  The latter does not posses any filamentary
structure in the vorticity field, witnessing the fact that the vortex
stretching mechanism, which is responsible for the forward cascade in
standard 3D systems, is here reduced. Similarly, the forward regime
does not possesses any coherent helicity signal at variance with the
inverse regime which shows strong non-homogeneity in the helical
spatial distribution.

In Fig.~\ref{fig:2} we show a typical evolution of the energy spectrum
obtained from Eq.~(\ref{eq:ns+++}) by initializing the flow with
energy only at high wave-numbers.  The development of an inverse
cascade with a Kolmogorov spectrum $E(k) \sim k^{-5/3}$ is unambiguous.

\begin{figure}[!t]
  \begin{center}
  \includegraphics[scale=0.7,draft=false]{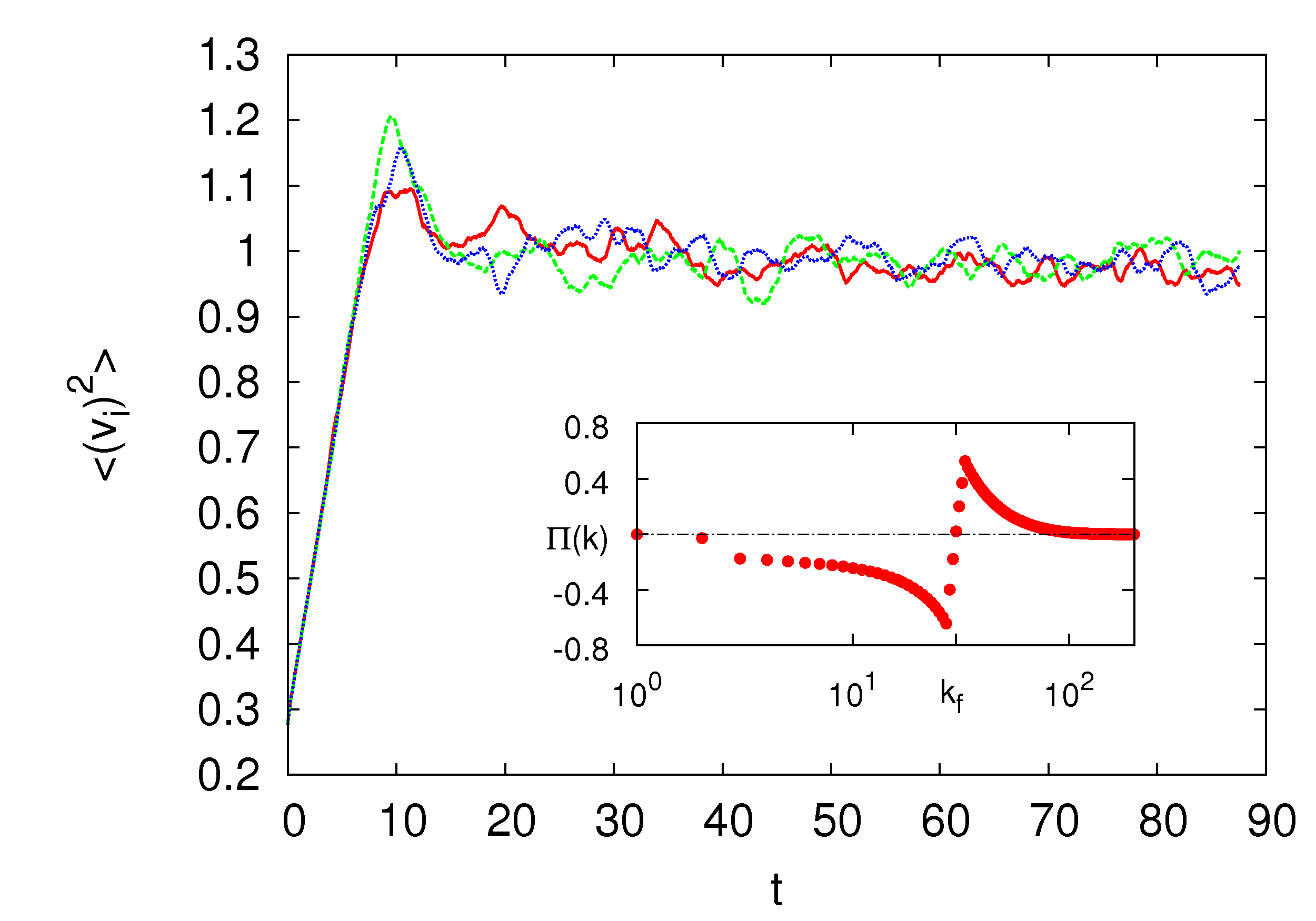}
  \caption{(color online) Evolution of the three components of the
    turbulent kinetic energy as a function of time, $\langle
    (v_i)^2\rangle $, with $i=x$ (red, solid line) $i=y$ (green,
    dashed line) and $i=z$ (blue, dotted line).  Inset: energy flux,
    $\Pi(k)$, in the Fourier space.  Notice the clear negative
    plateaux in the inertial range $ k < k_f$.}
  \label{fig:3}
\end{center}
\end{figure}

In absence of a large-scale dissipative mechanism, the inverse cascade
would accumulate the kinetic energy in the lowest available mode,
originating a {\it condensed state}~\cite{xia}. In order to avoid this
phenomenon we made a second series of numerical simulations, adding an
hypo-viscosity at large scale $\propto \Delta^{-1} v $.  In such a
case, the total kinetic energy becomes stationary as shown in
Fig.~\ref{fig:3}, and is equally distributed among the three velocity
components, showing that the flow is fully isotropic.  This allows to
study scaling properties without having to cope with anisotropic
sub-leading contributions~\cite{biferale_procaccia}.  In the inset of
Fig.~\ref{fig:3}, we show the stationary energy flux in Fourier space,
defined as $\Pi(k) \equiv (d/dt) \int_{k}^{\infty} E(p) dp$ where time
derivative is computed by taking into account only the non-linear
terms of Eq.~(\ref{eq:ns+++}).  The negative plateaux in the inertial
range of wave-numbers is a clear indication of the large-scale energy
transfer.

\begin{figure}[!t]
\begin{center}
\includegraphics[scale=0.7,draft=false]{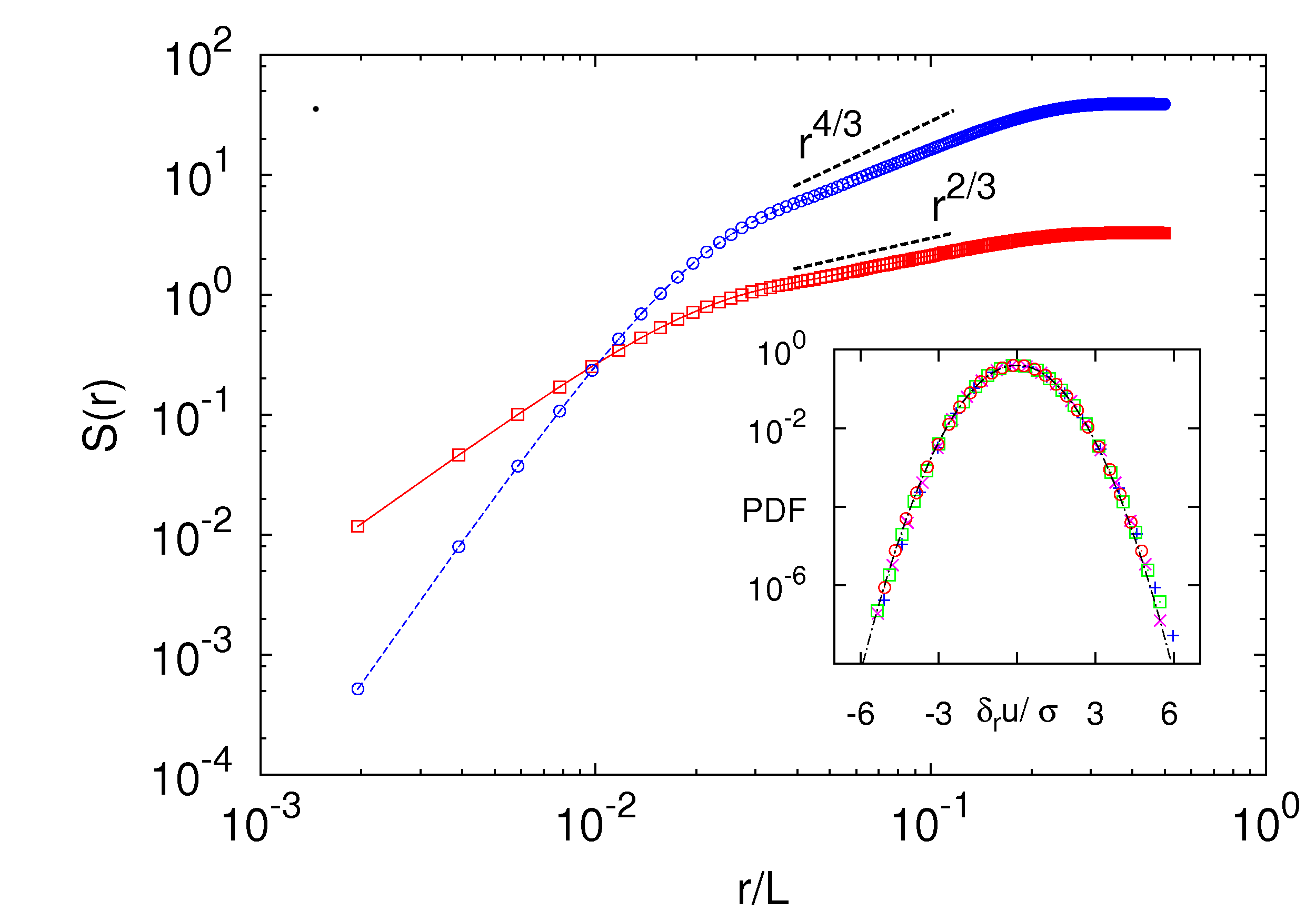}
\caption{(color online) Second order (squares) and fourth order
  (circles) structure functions in real space. Black solid and dashed
  lines represent the dimensional scaling $\propto r^{2/3}$ and
  $r^{4/3}$, respectively.  Inset: pdfs of $\delta_r v$ at various
  scale $r=[1/4,1/8/,1/16,1/32]L$, compared with the Gaussian
  distribution (dash-dotted line).  Notice the perfect rescaling,
  supporting the absence of intermittency.}
\label{fig:4}
\end{center}
\vskip -0.5cm
\end{figure}

The inverse cascade which arise from Eq.~(\ref{eq:ns+++}) is not
intermittent.  The probability distribution functions (pdfs) of the
longitudinal velocity increments $\delta_r v = [\bv(\bx +\br)
-\bv(\bx)] \cdot \hat \br$ at distance $r$ within the inertial range
are self-similar and almost Gaussian (see inset of Fig.~\ref{fig:4}).
The scaling of the second and the fourth order moment of velocity
increments $S_2(r) = \langle (\delta_r v)^2 \rangle; S_4(r) = \langle
(\delta_r v)^4 \rangle$ follow the dimensional scaling $S_p(r) \sim
r^{p/3}$ (see Fig.~\ref{fig:4}).  This is a signature of all known
inverse cascades: when fluctuations are transferred from faster to
slower degrees of freedom~\cite{biferale_shell}.
Previous studies have shown the possibility to produce large scale
motion by non-parity invariant small-scales forcing only at small
Reynolds numbers or in the quasi-linear regime~\cite{sulem}.
Conversely, our results do not trivially originate from the projection
of the forcing on the positive helicity states but is a genuine effect
of the non-linear dynamics.  To assess this issue we performed a test
simulation of the complete NS equation with the same projected forcing
used in Eq.~(\ref{eq:ns+++}).  After an initial transient in which
part of the energy accumulates at the forcing scale, a direct cascade
sets in and all the energy injected is transferred toward small
scales. This excludes the possibility that the forcing alone could be
responsible for the inverse energy transfer observed in the decimated
NS equation.

In conclusions, we have presented theoretical and numerical evidences
that a {\it screwed} version of the NS equations, such that only modes
with a given sign of helicity are retained, displays inverse energy
transfer mechanism. This phenomenon, which has been previously
observed only in 2D turbulence or in strongly anisotropic 3D flows
under bidimensionalization effects, is here observed for the first
time in a fully isotropic 3D system and is intrinsically connected to
the non-linear dynamics of {\it all} flows in nature.

The scientific impact of our findings is manifold.  First, it allows
to highlight those backward events in the energy transfer mechanism
which are known to exist also in {\it untruncated} NS equations and
that are one of the main theoretical and applied challenges, see
e.g.~\cite{meneveau} for the case of sub-grid modelling in Large Eddy
Simulations. Second, the link between backward energy events with the
helical nature of triad interaction, shows the key role of the coupled
energy-helicity dynamics. Third, by clearly detecting which triadic
interaction is responsible for forward and backward energy transfer,
we pave the road for closure and analytical approaches aimed at
understanding the whole energy transfer distribution.

This study also opens the way to further investigations. An obvious
extension would be to integrate Eq.~(\ref{eq:ns+++}) with a large
scale forcing.  In this case a pure forward helicity cascade must
develop, provided that energy is removed at the forcing scale to avoid
pile-up of fluctuations.  More interesting, one could consider the
case of a complementary decimation with respect to the one discussed
here, i.e. eliminating only those triads that transfer energy
backward. It is very tempting to speculate that such system could
display a direct energy cascade with reduced --or even vanishing--
intermittency, because one has removed all the {\it obstacles},
i.e. those events in which the forward energy transfer is stopped
and/or reversed by the interaction with the helicity flux. Numerical
simulations exploring these cases are ongoing and will be reported
elsewhere. Finally, similar decomposition may shed lights also in the
evolution of conducting fluids, where three invariants, kinetic plus
magnetic energy, cross helicity and magnetic helicity are known to
produce a reach phenomenology \cite{mhd}.
We acknowledge useful discussion with U. Frisch.  L.B. acknowledge the
kind hospitality from the Observatoire de la Cote d'Azur in Nice where
part of this work was done. We acknowledge the European COST Action
MP0806.

\end{document}